\documentclass[conference]{IEEEtran}
\IEEEoverridecommandlockouts

\usepackage{cite}
\usepackage{amsmath,amssymb,amsfonts}
\usepackage{csquotes}
\usepackage{breqn}
\usepackage{algorithmic}
\usepackage{graphicx}
\usepackage{textcomp}
\usepackage{setspace}
\usepackage{subcaption}
\usepackage{xcolor}
\def\BibTeX{{\rm B\kern-.05em{\sc i\kern-.025em b}\kern-.08em
    T\kern-.1667em\lower.7ex\hbox{E}\kern-.125emX}}

\begin{document}

\title{Generalized Distance Metric for Various DHT Routing Algorithms in Peer-to-Peer Networks}

\author{\IEEEauthorblockN{Rashmi Kushwaha, Shreyas Kulkarni, Yatindra Nath Singh}
\IEEEauthorblockA{
\textit{Department of Electrical Engineering} \\
\textit{Indian Institute of Technology, Kanpur}\\
Kanpur, India \\
and \\
\textit{YRRNA Systems Lab Pvt Ltd}\\
rashmikushwaha221@gmail.com, shreyas.kulkarni29@gmail.com, ynsingh@iitk.ac.in}
}

\maketitle
\thispagestyle{plain}
\pagestyle{plain}
\begin{abstract}
We present a generalized distance metric that can be used to implement routing strategies and identify routing table entries to reach the root node for a given key, in a DHT (Distributed Hash Table) network based on either Chord, Kademlia, Tapestry, or Pastry. The generalization shows that all the above four DHT algorithms are in fact, the same algorithm but with different parameters in distance representation. We also proposes that nodes can have routing tables of varying sizes based on their memory capabilities but with the fact that each node must have at least two entries, one for the node closest from it, and the other for the node from whom it is closest in each ring components for all the algorithms. Messages will always reach the correct root nodes by following the above rule. We also further observe that in any network, if the distance metric to define the root node in the DHT is same at all the nodes, then the root node for a key will also be the same, irrespective of the size of the routing table at different nodes.
\end{abstract}


\section{Introduction}
The development of technology has created a world where smart devices are all around us. In the client-server paradigm, all these devices need to connect to some server for retrieving or depositing the information. Consequently, servers are essential for all internet-based services. This has created numerous challenges for the client-server design, such as scalability to handle increasing number of users, resilience against server failure, attacks to disrupt services, etc.. Every central server has disadvantage of single point of failure, and will also has scaling limitations. They need to be made more reliable with huge capacities, thereby increasing the cost. Also, the client-server paradigm allows the companies to exercise user surveillance by just snooping on the server. To handle these challenges, many researchers have explored peer-to-peer architectures for provisioning of all kind of services.
\par A Peer-to-Peer (P2P) system is a distributed architecture in which each peer acts as a client taking services from other peers as well as a server providing some service to others. In this design, there are no central servers involved. Peer-to-peer architecture saves a lot of expenditure incurred on servers to improve scalability and reliability. There are many other benefits of peer-to-peer networks, such as data redundancy, pseudonymity, and massive scalability, but it all depends on the efficient methods to find the needed data within the network. This led to a large number of studies investigating distributed data indexing\cite{b9}.
\par Initially, peer-to-peer systems were built with centralized indexing servers, in which data search requests are handled by servers and the actual resource sharing used to happen directly between peers. But again, server scalability was a bottleneck. Further, any impact on the server resulted in the impact on the whole network. One such example was Napster\cite{b1}.
\par At the same time, completely distributed, unstructured P2P network architectures like Gnutella\cite{b2} and Freenet\cite{b3} have also evolved. The management of content was unrelated to a peer. The flooding of requests was the only option to search for any data. There is no guarantee that one will find the resource even if it exists somewhere in the network. In terms of performance, these were extremely inefficient.
\par DHT-based structured P2P network architectures were developed to be more efficient and assure faster resource discovery. DHTs can be used in peer-to-peer (P2P) networks to provide decentralized and self-organizing services such as file sharing, distributed file storage, and distributed data management. DHT essentially employs the idea of consistent hashing. Each node is assigned a random nodeID of same size as that of hash of keys (hashID) to be used for indexing. A distance metric is defined between any two points in the nodeID space. In a Distributed Hash Table (DHT), every content is stored at a node (peer) based on the hash of the search key of the content. Each peer is responsible for all the contents whose keys’ hash values are closest to its nodeID. Thus, the notion of distance between nodeID and the hash value (hashID) of keys need to be defined. Each $\langle key, value  \rangle$ pair corresponds to a point identified by hash\{key\} in nodeID space. The nodeID closest to this hash\{key\} is responsible for storing the $\langle key, value  \rangle$ pair.
\par Consistent hashing allows users to be added and removed in any arbitrary order. In this, less $\langle key, value  \rangle$ pairs need to be redistributed when a peer joins or leaves the network. When a node leaves, the $\langle key, value  \rangle$ pairs stored with it, are redistributed to other adjacent nodes. When a new node joins the network, it takes over the responsibility of certain $\langle key, value  \rangle$ pairs from the other neighbouring nodes. Due to uniform distribution of nodes in the nodeID space and use of cryptographic hashing, load balancing is achieved for $\langle key, value  \rangle$ storage. Consistent hashing also takes care of churn as only a smaller number of entries are to be relocated. Churn means frequent exit and entry of nodes from and into the network respectively.
\par The most prominent DHT algorithms are Chord\cite{b4}, Tapestry\cite{b5}, Pastry\cite{b6}, and Kademlia\cite{b7}. Each one of them is defined differently, based on (I) a distance metric and (II) a way of maintaining pointers to other nodes in the DHT network (called routing tables/ finger tables/ leafset/ neighbour tables). On careful examination, we observe that all the four algorithms—chord\cite{b4}, tapestry\cite{b5}, kademlia\cite{b7}, and pastry\cite{b6}—can be described by a generalized distance metric. By varying parameters of the distance metric, we can switch between these algorithms. It means that all four algorithms are not different, but rather the result of changing some parameters in a single algorithm.
\par In this paper,
\begin{itemize}
\item We present a generalized formula for distance metric, which can be considered as an all-in-one formula to represent different DHT algorithms. Many studies have been done on comparative performance\cite{b9}, \cite{b11}, but no study had given framework with a single generalized formula.
\item We also propose that there can be different sizes of routing tables at different nodes in the same network and still the uniqueness of root node is maintained, if same distance metric to define root node is used by all of them. 
\end{itemize}
\par In this paper, section II presents the related work, section III presents the generalized formula and how it represents different DHT algorithms. Section IV presents examples to illustrate the applicability of generalized formula, and section V focuses on study of impact of variable routing table sizes at different nodes in same network. Finally conclusions are presented.

\section{Related work}

A Distributed Hash Table (DHT) is a distributed system that provides an efficient lookup service similar to a hash tables used in databases at a server. The DHT stores $\langle key, value  \rangle$ pairs, and any participating node can easily and effectively store or retrieve the value associated with a given key. Research in this field has concentrated on developing hashing algorithms that uniformly distribute load, while transversing least number of hops using efficient routing. Further, each node needs to be aware of much less number of nodes to form routing table, compared to the total population of nodes in a large network.
\par In literature, number of well-known DHTs have been proposed and used in P2P networks, e.g, Chord\cite{b4}, CAN\cite{b10}, Tapestry\cite{b5}, Kademlia\cite{b7}, Pastry\cite{b6} etc..
\par Chord\cite{b4} is a distributed lookup protocol that makes use of finger tables to locate the root node of a given key. The efficiency provided by the Chord protocol for locating the root node is O(log N), where N is the number of nodes in the network.
\par The routing in Tapestry\cite{b5} is based on a hierarchical structure encoded in a \enquote{routing table}, which is similar to a prefix-based routing table in IP networks. Each node maintains a routing table that maps a prefix of current node's ID to a node representing a set of nodes. The matching prefix length of each entry in the routing table increases as the distance between the node and the destination node decreases.
\par Pastry\cite{b6} is similar to Chord, but with certain differences. Pastry routes information to the numerically closest nodeID while in chord, routing is done to closest nodeID larger than or equal to hashID in cyclic sense. Pastry uses almost similar routing table as Tapestry, but also use a leafset table of size $k$, which contain $k/2$ immediate successor and $k/2$ immediate predecessor nodes. Network locality is taken into account in pastry while choosing one of the many options for each routing table entry. The routing process starts at the requesting node, and at each hop, the leaf set and routing table are consulted to determine the next hop. When the destination node is reached, the data is either stored or retrieved. It is considered to be more efficient than Chord in terms of network traffic and message complexity.	Average time a message takes to reach the root node, is much less compared to chord.
\par Kademlia\cite{b7} uses a hierarchical structure to organize nodes, and uses an XOR-based distance metric to determine the proximity of points in nodeID space. Since XOR is symmetric, Kademlia nodes can get lookup requests from the same mix of nodes that are in their routing tables if criteria to choose from multiple option for an entry is also smaller XOR distance. This makes it easier to maintain NAT transversal.
\par These are the most well-known DHTs, but there are other DHTs also that have been proposed and implemented in P2P networks. Each DHT has its own strengths and weaknesses, and the choice of DHT to use in a particular application depends on the specific requirements of the application. For example, Accordion\cite{b8}, a DHT protocol proposed by Li, balances the need for low lookup hop-count and low timeout probability by bounding its overhead traffic to a user-specified bandwidth budget and selecting a routing table size that optimizes lookup latency. CAN\cite{b10} uses hashID space range instead of nodeID along with endpoint address in the routing table. It tries to equalize the hashID volume for which nodes are responsible. 

\section{Generalized Distance Metric Formula} 
A generalized equation that can be used to define distance which is used to decide routing table entries to reach the root node for a given key, in DHT networks such as Chord, Kademlia, Tapestry, and Pastry can be represented as

\begin{equation}\label{eq:GD}
 D=\sum_{i=0}^{k-1}\left\{\left(r_i-h_i+2^d\right) \bmod 2^d\right\} \times 2^{d \times(i)}
\end{equation}

Here, nodeID and hashID (hash of keys) are represented by 
\begin{equation*}
 R=r_{k-1} r_{k-2} \ldots r_0   
\end{equation*}
and 
\begin{equation*}
H=h_{k-1} h_{k-2} \ldots h_0,
\end{equation*}
respectively. Here ${r_i}$ and ${h_i}$ are $2^d$-ary digits, and each of them can be represented by $d$-bits. $k$ is the number of digits in the nodeIDs and hashIDs.
\par Any key-value pair will be stored at a node whose nodeID is equal to or closest from the hashID. This node is termed as root node of the hashID.
\begin{center}
$\min_R D(R, H)$,
then $R$ is the root node of $H$.
\end{center}
\par In chord and pastry, nodeID and hashID are considered as one single digit with the base of maximum possible value (for a 4-bit system, the max possible value is 15+1= 16). Thus, all the bits in $H$ and $R$ are used for digits. 
The distance metric in Eqn(\ref{eq:GD}) becomes
\begin{equation} \label{eq:GD2}
D= \left(R-H+2^d\right) \bmod 2^d   
\end{equation}

In Eqn(\ref{eq:GD}), $i$ can only take value 0, and d is the number of bits in nodeID and hashID. This is used in chord.
Maximum value of $R$ and $H$ is, 
\begin{equation*}
{(2^d-1)=H_{max}}.
\end{equation*}
We can write the  distance in Eqn(\ref{eq:GD2}) as
\begin{equation}\label{eq:DG3}
 {D= \left(R-H+H_{max}+1\right) \bmod (H_{max}+1)}.  
\end{equation}
For Pastry, $$\min \left(\left\{\left(R-H+2^d\right) \bmod 2^d\right\},\left\{\left(H-R+2^d\right) \bmod 2^d\right\}\right),$$ is used as distance to decide the root node.

\par In Kademlia, single bit per digit $(d=1)$ is used, thus the distance metric in Eqn(\ref{eq:GD}) becomes, 
\begin{equation}\label{eq:GD4}
 D=\sum_{i=0}^{k-1}\left\{\left(r_i-h_i+2\right) \bmod 2\right\} \times 2^{i}
\end{equation}
$\left\{\left(r_i-h_i+2\right) \bmod 2\right\}$ for bits ${r_i}$ and ${h_i}$ is equal to ($r_i \oplus h_i),$ thus $D$ is also called XOR distance.
\par For Tapestry, the digit size can be changed depending on the tapestry specification chosen by the implementers. Commonly used digit sizes are either 3-bits or 4-bits per digit. So, for 4-bits per digit $(d=4)$, we can represent the distance in Eqn(\ref{eq:GD}) as
\begin{equation} \label{eq:GD5}
D= \sum_{i=0}^{k-1}\left\{\left(r_i-h_i+2^4\right) \bmod 2^4\right\} \times 2^{4 \times(i)}
\end{equation}
\par In all the above algorithms, the nodeID space is partitioned logarithmically to create a routing table. So, each node will have pointers to a node representing half, one-fourth, one-eighth partition (for base-16 Tapestry, pointers to sixteen subspaces for each size are maintained. The subspaces sizes are $1/16$, $1/16^2$, $1/16^3$ and so on) of nodeID space. In general, each node can have
\{$(2^d-1)*k$\} entries in its routing tables for Pastry, Tapestry and Kademlia.
\par A special case arises when $\mathrm{d}=$ \textit{number of bits in the nodeID}. Now, Tapestry routing table will have a single column and entries for all the nodes other than current node will exist. They form a ring and routing tables like a finger table but with all the nodes in it. If we form logarithmic partition relative to current node in the column and at least one predecessor and successor is maintained, we get the chord algorithm.
\par We can extend this logic to Tapestry with $d=4$. In each column, we can organise nodes as finger table formed by the nibble corresponding to the column. So, Kademlia, Tapestry and Chord represent same algorithm with Kademlia and Chord being two extremes.
\par In general, a node with nodeID $C$ in chord will have root node of 
  \begin{equation}\label{eq:GD6}
  \left\{\left(C+2^{mi}\right) \bmod 2^{k}\right\} \text { for } 0 \leq i \leq k-1
  \end{equation}as one of the entries in finger table. Here, $k$ is the integral multiple of $m$. Thus, $\left[\log _{2^m}\left(2^{k}\right)\right]$ entries will be there in the finger table. Additionally, immediate succeeding and preceding neighbors also need to be maintained by each node. This neighbor table is also called leafset. All hashIDs greater than preceding nodeID and less than equal to current nodeID are the responsibility of the current node $C$.
  \par In Pastry also, leafsets are maintained by all the nodes, but the routing tables are maintained as a matrix of size \{$k*(2^d-1)$\}. Each cell in row $i$, and column $j$ will contain one of the nodes with nodeID which have $s_{k-i}=j, \quad(1 \leq i \leq k)$.  
  \par $s_{k-1}, s_{k-2}$, $\ldots, s_{k-i+1}$ will be same as that for the current node. The cell corresponding to the $s_{k-i}$ digit of the current node will contain the current nodeID. Each node will transfer a query for root node of hashID $H$ to a node with longest matching nodeID in the current node’s routing table. When a longer match can no longer be found, then leafset is used for forwarding to a node closest to the hashID. The root node search terminates when current node finds itself to be the root node using its leafset.
  \par In Kademlia, again similar routing table is maintained, but now matrix size is \{$\left(\log _2 N-1\right) \times 1$\}. Also, there is no leafset. The cell $(i,1)$ will contain nodeID such that $s_{k-1}, s_{k-2}, \ldots, s_{k-i+1}$ bits will be the same as in the current node’s nodeID. The $s_{k-i}$ bit in the routing table entry will be opposite of the $s_{k-i}$ bit of current node’s nodeID. All other bit’s i.e., $s_{k-i-1}, \ldots, s_{0}$, can take any value. This selection can be done to minimize RTT, geographical distance or XOR distance. Choosing best XOR distance leads to symmetric routing tables i.e. if $A$ has $B$ in its RT, then $B$ will also have $A$ in its RT.
  \par In Tapestry, the routing table is similar to what is used in Pastry, but there are no leafset tables. There are $(2^d-1)$ rows and $(log_2N)/d = k$ columns. A column $i$ contains entry of nodes which have $s_{k-1}, s_{k-2}, \ldots, s_{k-i+1}$ digits which are same as current node, $s_{k-i}$ must be different than what is in the current node’s nodeID, and hence $(2^d-1)$ possible entries are kept in each column.
  \par Every node simply forwards a query for a hashID to the entry in its routing table, which has the least distance from hashID as per the distance metric. If the current node’s nodeID is having least distance, then, the current node will be the root node, and it should do the needful to find the $\langle key-value  \rangle$ pair and send it to the querying node or to store the received $\langle key-value  \rangle$ in the local index depending if message is query or publish message. 
  
\section{Examples}
Consider, a network with 16-bit nodeIDs. Let the current node be \textbf{03A6}. Assume that we need to route a query with hashID \textbf{4EFA} to its root node. Let us assume that nodes present in the system are \textbf{0156}, \textbf{0359}, \textbf{0379}, \textbf{03A6}, \textbf{03A9}, \textbf{03AF}, \textbf{04A5}, \textbf{25AB}, \textbf{456B}, \textbf{4ABC}, \textbf{4E56},  \textbf{4EAB}, \textbf{4ECD}, \textbf{4EF7}, \textbf{4EFB}, \textbf{4EFC}, \textbf{4EFD}, and \textbf{6754}.
We now see how the tables are formed and how the message is routed in the different DHT algorithms.
\vspace{0.3cm}
\begin{enumerate}
 \item {\large{\textbf{{Tapestry}}}}:
 Consider a 4-digit hexadecimal nodeID (4-bits per digit). When we put $d=4$ in generalized distance metric formula (Eqn(\ref{eq:GD})), we have distance metric as,
\begin{equation*}
 D=\sum_{i=0}^{k-1}\left\{\left(r_i-h_i+2^4\right) \bmod 2^4\right\} \times 2^{4 \times(i)}.
\end{equation*}
\par Each node will have a routing table size of \{$(2^d-1)$ rows * $k$ columns\}. Here there will be \{$15$ rows * $4$ columns\}.
Figure \ref{fig:1} and table \ref{table:a} show routing entries of nodeID 03A6. For hashID 4EFA, 456B is the highest matching prefix. So, it will forward the query to 456B.
\par
\hspace{5mm} Tables \ref{table:b}, \ref{table:c} and \ref{table:d} show the routing table of 456B, 4EAB and 4EFC respectively.
Figure \ref{fig:fig21} shows, how query will be forwarded based on matching the prefix of nodeID in Tapestry. 
In the figure \ref{fig:T2}, node 4EFC has two corresponding nodes to 4EFA in L4, but from the distance metric of the tapestry, the root node of 4EFA will be 4EFB. 
\begin{figure}[htp]
    \centering
    \includegraphics[width=07cm]{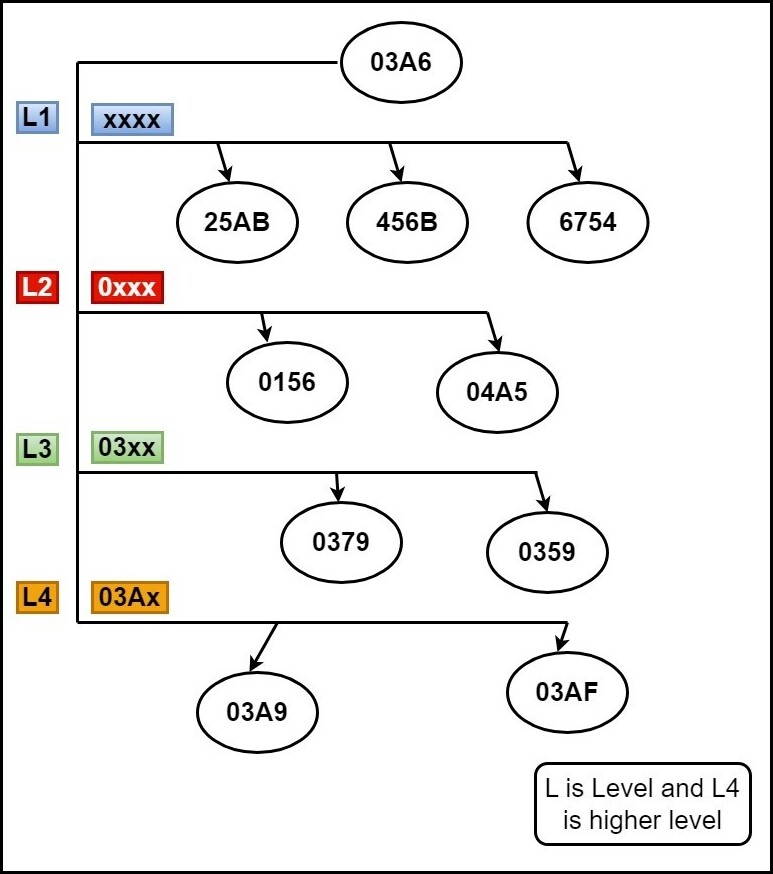}
    \caption{Routing entries of 03A6 in Tapestry/Pastry Network. Higher level shows higher matching of digits} \label{fig:1}
\end{figure}
\begin{table}[h!]
\centering
\caption{Routing Table of 03A6 in Tapestry Network}
\label{table:a}
\begin{tabular}{|c|c|c|c|}
\hline
L1 & L2 & L3 & L4 \\ \hline
25AB & 04A5 & 0359 & 03A9 \\
\textbf{456B} & 0156 & 0379 & 03AF \\
6754 & & & \\ \hline
\end{tabular}
\end{table}
\begin{table}[h!]
\centering
\caption{Routing Table of 456B in Tapestry Network}
\label{table:b}
\begin{tabular}{|c|c|c|c|}
\hline
L1 & L2 & L3 & L4 \\ \hline
6754 & 4ABC &  &  \\
03A6 & \textbf{4EAB} & &  \\
25AB & & & \\ \hline
\end{tabular}
\end{table}
\begin{table}[h!]
\centering
\caption{Routing Table of 4EAB in Tapestry Network}
\label{table:c}
\begin{tabular}{|c|c|c|c|}
\hline
L1 & L2 & L3 & L4 \\ \hline
6754 & 456B & 4ECD &  \\
0359 & 4ABC & \textbf{4EFC} &  \\
25AB & & 4E56 &  \\ \hline
\end{tabular}
\end{table}
\begin{table}[h!]
\centering
\caption{Routing Table of 4EFC in Tapestry Network}
\label{table:d}
\begin{tabular}{|c|c|c|c|}
\hline
L1 & L2 & L3 & L4 \\ \hline
6754 & 456B & 4E56 & 4EFD \\
03A9 & 4ABC & 4EAB & 4EF7 \\
25AB & & 4ECD & \textbf{4EFB} \\ \hline
\end{tabular}
\end{table}
    \begin{figure}[htp]
    \centering
    \includegraphics[width=07cm]{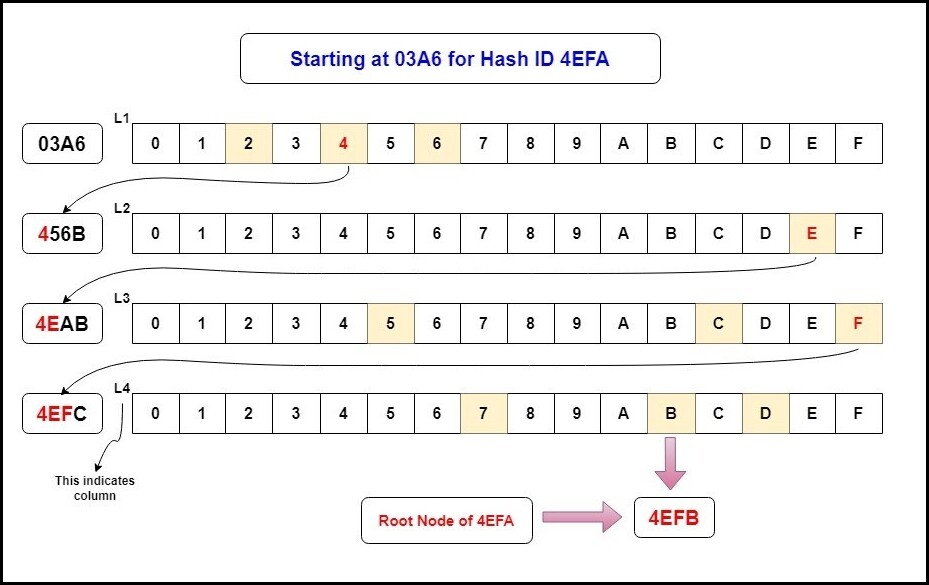}
     \caption{Routing is done with the matching prefix increasingly}   
    \label{fig:fig21}
\end{figure}
    \begin{figure}[htp]
    \centering
    \includegraphics[width=07cm]{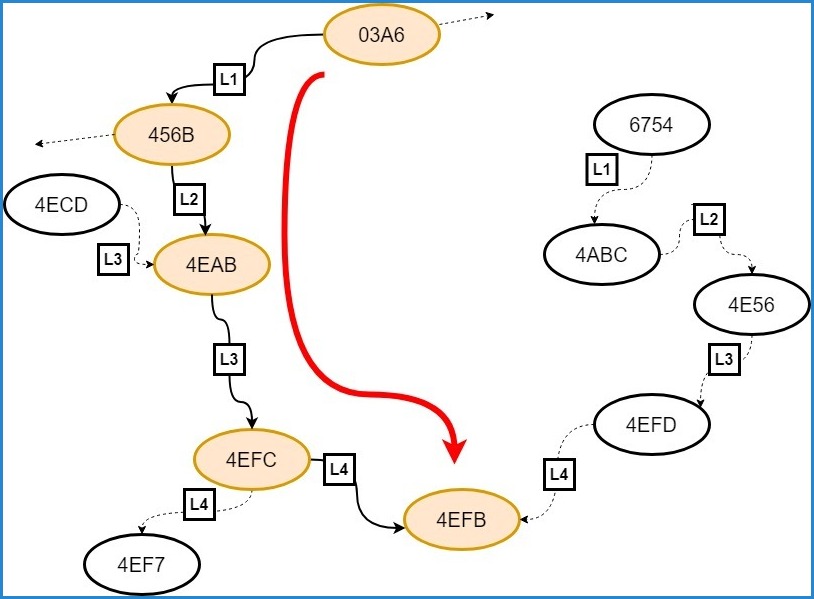}
     \caption{Route to find the root node of 4EFA from 0326 in Tapestry}   
    \label{fig:T2}
\end{figure}
\vspace{0.3cm}
\item {\large{\textbf{{Kademlia}}}}:
When we put $d=1$ in generalized formula of Eqn(\ref{eq:GD}), we have 16-bit nodeId (1-bit per digit) and we have distance metric as given in Eqn(\ref{eq:GD4}).
\par Here, one can observe that $$\sum_{i=0}^{k-1}\left\{\left(r_i-h_i+2\right) \bmod 2\right\} \times 2^{(i)}$$ is same as $\sum_{i=0}^{k-1}\left\{r_i \oplus h_i\right\} \times 2^{(i)},$ because $(\left(r_i-h_i+2\right) \bmod 2)$ is just another representation of XOR. For the same set of nodes, we now represent all nodeIDs as binary numbers. \par
\hspace{5mm} Self nodeId; \textbf{03A6 is 0000 0011 1010 0110}. We again find the Root node of hashID \textbf{4EFA: 0100 1110 1111 1010}. Table \ref{table:1} shows nodes present in the system. We have represented nodeIDs in the form of binary number.
Each node will have a routing table of size $(2^d-1)$ rows * $k$ columns. Here it will be $1$ rows * $16$ columns.\par
\begin{table}[h!]
\centering
\caption{Nodes present in Kademlia}
\label{table:1}
\begin{tabular}{|c|c|}
\hline
0156 & 0000 0001 0101 0110 \\ \hline
0359 & 0000 0011 0101 1001 \\ \hline
0379 & 0000 0011 0111 1001 \\ \hline
03A6 & 0000 0011 1010 0110 \\ \hline
03A9 & 0000 0011 1010 1001 \\ \hline
03AF & 0000 0011 1010 1111 \\ \hline
04A5 & 0000 0100 1010 0101 \\ \hline
25AB & 0010 0101 1010 1011 \\ \hline
456B & 0100 0101 0110 1011 \\ \hline
4ABC & 0100 1010 1011 1100 \\ \hline
4E56 & 0100 1110 0101 0110 \\ \hline
4EAB & 0100 1110 1010 1011 \\ \hline
4ECD & 0100 1110 1100 1101 \\ \hline
4EF7 & 0100 1110 1111 0111 \\ \hline
4EFB & 0100 1110 1111 1011 \\ \hline
4EFC & 0100 1110 1111 1100 \\ \hline
4EFD & 0100 1110 1111 1101 \\ \hline
6754 & 0110 0111 0101 0100 \\ \hline
\end{tabular}
\end{table}
\begin{table}[h!]
\centering
\caption{Routing Table of 03A6 in Kademlia Network}
\begin{tabular}{|c|c|}
\hline
Column No. & NodeID \\ \hline
1 & - \\ \hline
2 & \textbf{456B} \\ \hline
3 & 25AB \\ \hline
4 & - \\ \hline
5 & - \\ \hline
6 & 04A5 \\ \hline
7 & 0156 \\ \hline
8 & - \\ \hline
9 & 0379 \\ \hline
10 & - \\ \hline
11 & - \\ \hline
12 & - \\ \hline
13 & 03AF \\ \hline
14 & - \\ \hline
15 & - \\ \hline
16 & - \\ \hline
\end{tabular}
\label{tab:03A6}
\end{table}
\begin{table}[h]
    \begin{subtable}[h]{0.22\textwidth}
        \centering
        \begin{tabular}{|c | c |}
        \hline
            \multicolumn{2}{|c|}{Routing Table} \\\hline 
               Column No. & NodeID \\ \hline
1 & - \\ \hline
2 & 04A5 \\ \hline
3 & 6754 \\ \hline
4 & - \\ \hline
5 & \textbf{4E56} \\ \hline
6 & - \\ \hline
7 & - \\ \hline
8 & - \\ \hline
9 & - \\ \hline
10 & - \\ \hline
11 & - \\ \hline
12 & - \\ \hline
13 & - \\ \hline
14 & - \\ \hline
15 & - \\ \hline
16 & - \\ \hline
       \end{tabular}
       \caption{}
       \label{tab:456B}
    \end{subtable}
    \hfill
    \begin{subtable}[h]{0.26\textwidth}
        \centering
        \begin{tabular}{|c | c |}
        \hline
            \multicolumn{2}{|c|}{Routing Table} \\\hline 
                Column No. & NodeID \\ \hline
1 & - \\ \hline
2 & 04A5 \\ \hline
3 & 6754 \\ \hline
4 & - \\ \hline
5 & 456B \\ \hline
6 & 4ABC \\ \hline
7 & - \\ \hline
8 & - \\ \hline
9 & \textbf{4ECD} \\ \hline
10 & - \\ \hline
11 & - \\ \hline
12 & - \\ \hline
13 & - \\ \hline
14 & - \\ \hline
15 & - \\ \hline
16 & - \\ \hline
        \end{tabular}
        \caption{}
        \label{tab:4E56}
     \end{subtable}
       \caption{Routing Table of 456B and 4E56 in Kademlia Network respectively}   
     \end{table}
     \begin{table}[h]
    \begin{subtable}[h]{0.22\textwidth}
        \centering
        \begin{tabular}{|c | c |}
        \hline
            \multicolumn{2}{|c|}{Routing Table} \\\hline 
               Column No. & NodeID \\ \hline
1 & - \\ \hline
2 & 04A5 \\ \hline
3 & 6754 \\ \hline
4 & - \\ \hline
5 & 456B \\ \hline
6 & 4ABC \\ \hline
7 & - \\ \hline
8 & - \\ \hline
9 & 4E56 \\ \hline
10 & 4EAB \\ \hline
11 & \textbf{4EFD} \\ \hline
12 & - \\ \hline
13 & - \\ \hline
14 & - \\ \hline
15 & - \\ \hline
16 & - \\ \hline
       \end{tabular}
       \caption{}
       \label{tab:4ECD}
    \end{subtable}
    \hfill
    \begin{subtable}[h]{0.26\textwidth}
        \centering
        \begin{tabular}{|c | c |}
        \hline
            \multicolumn{2}{|c|}{Routing Table} \\\hline 
               Column No. & NodeID \\ \hline
1 & - \\ \hline
2 & 04A5 \\ \hline
3 & 6754 \\ \hline
4 & - \\ \hline
5 & 456B \\ \hline
6 & 4ABC \\ \hline
7 & - \\ \hline
8 & - \\ \hline
9 & 4E56 \\ \hline
10 & 4EAB \\ \hline
11 & 4ECD \\ \hline
12 & - \\ \hline
13 & 4EF7 \\ \hline
14 & 4EFB \\ \hline
15 & - \\ \hline
16 & 4EFC \\ \hline
        \end{tabular}
        \caption{}
        \label{tab:4EFD}
     \end{subtable}
       \caption{Routing Table of 4ECD and 4EFD in Kademlia Network}   
     \end{table}
\hspace{5mm} Table \ref{tab:03A6} shows routing Table of \textbf{03A6: 0000 0011 1010 0110}. In this table, columns and rows are interchanged. Each column $i$ represents $(i-1)$ MSB bits matching with the self nodeId and $ith$ MSB will be opposite of $ith$ MSB of the current node. All other bits can take any value. If there are more than one nodes which can be placed at an entry in routing table, then it will calculate XOR distance and use the entry with least distance from the current node. Empty cell represents that there are no nodes matching the corresponding pattern.\par
\hspace{5mm} For example in column No. 9, there are two nodeIds with similar numbers of bits matching: 0359 and 0379, but the XOR distance of 0379 is less than that of 0359 from 03A6, hence that is used.\par  
\hspace{5mm} 03A6 will hand over the query 4EFA to 456B. According to their respective routing tables, 456B will pass it to 4E56, 4E56 to 4ECD, 4ECD to 4EFD, and 4EFD to 4EFB. 4EFB is finally the root node.

\item {\large{\textbf{{Chord}}}}:
The nodeID and hashID will be considered as one single digit with the base of the maximum possible value+1 (for a 16-bit system, the maximum possible value+1 is 65535+1 = 65536).
We have the metric for distance as described in Eqn(\ref{eq:GD2}) as, 
\begin{equation*}
D= \left(R-H+2^d\right) \bmod 2^d .  
\end{equation*}
Each node with nodeID C will have a root node of  
    \begin{equation*}
    \left\{\left(C+2^{m i}\right) \bmod 2^k\right\} \text { for } 0 \leq i \leq{(k-1)/m}
    \end{equation*} in the finger table. Thus, the routing table will have $\left\{\log _{2^m}\left(2^k\right)\right\}$  entries.
    Additionally, the immediate successor and predecessor nodes are also maintained as leaf sets.\par
\hspace{5mm} Consider, m=2, k=16. So, each node will have $\left\{\log _{2^2}\left(2^{16}\right)\right\} =8 $ entries in their routing table and 1 successor and 1 predecessor in their leaf set.
Nodes present in chord network can be represented as shown in fig \ref{fig:image17}. Nodes are arranged in ascending order forming a ring structure. 
\begin{figure}[htp]
    \centering
    \includegraphics[width=07cm]{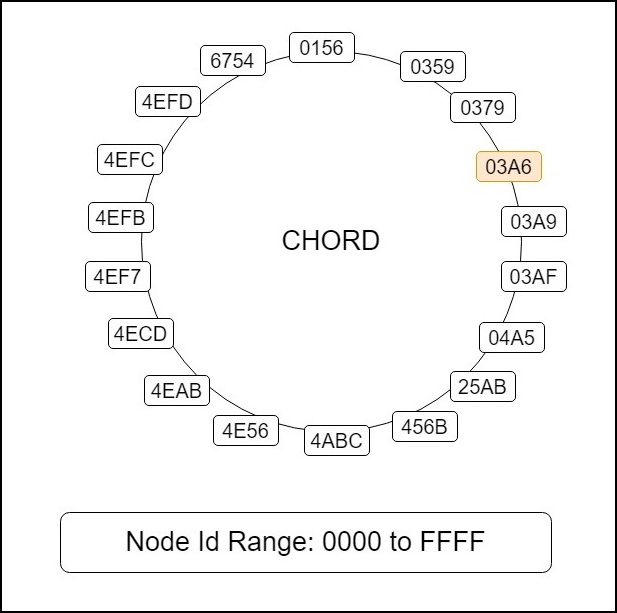}
    \caption{Nodes present in chord network}  
   \label{fig:image17}
\end{figure}\par
\hspace{5mm} For NodeID \textbf{03A6}, finger table entries are calculated as, 
   \begin{equation*}
    Root \left\{\left((03A6)_{16} +2^{2 i}\right) \bmod 2^k\right\} \text { for } 0 \leq i \leq {(k-1)/m}.  
   \end{equation*} \par
   Table \ref{tab:chord Table} shows routing table and leafset for this node. From this finger table, 03A6 will hand over query 4EFA to node 456B.\par
\begin{table}[h]
    \begin{subtable}[h]{0.22\textwidth}
    \centering
        \begin{tabular}{|c|c|}
        \hline
            \multicolumn{2}{|c|}{Routing Table} \\\hline 
                Value of \lq${i}$\rq & NodeID \\\hline
                0 & 03A9  \\\hline
                1 & 03AF \\\hline
                2 & 04A5 \\\hline
                3 & 04A5 \\\hline
                4 & 25AB \\\hline
                5 & 25AB \\\hline
                6 & 25AB \\\hline
                7 & \textbf{456B} \\\hline
       \end{tabular}
    \end{subtable}
    \hfill
    \begin{subtable}[h]{0.26\textwidth}
        \centering
        \begin{tabular}{|c|c|}
        \hline
            \multicolumn{2}{|c|}{Leafset} \\\hline     Predecessor(P) & 0379  \\\hline
                Successor(S) & 03A9  \\\hline
        \end{tabular}
     \end{subtable}
       \caption{Routing Table and Leafset of 03A6 in Chord Network}   
     \label{tab:chord Table}
\end{table}
\begin{table}[h]
    \begin{subtable}[h]{0.22\textwidth}
        \centering
        \begin{tabular}{|c | c |}
        \hline
            \multicolumn{2}{|c|}{Routing Table} \\\hline 
                Value of \lq${i}$\rq & NodeID \\\hline
                0 & 4ABC  \\\hline
                1 & 4ABC \\\hline
                2 & 4ABC \\\hline
                3 & 4ABC \\\hline
                4 & 4ABC \\\hline
                5 & 4ABC \\\hline
                6 & 6754 \\\hline
                7 & 0156 \\\hline
       \end{tabular}
    \end{subtable}
    \hfill
    \begin{subtable}[h]{0.26\textwidth}
        \centering
        \begin{tabular}{|l | l |}
        \hline
            \multicolumn{2}{|c|}{Leafset} \\\hline     Predecessor(P) & 25AB  \\\hline
                Successor(S) & \textbf{4ABC}  \\\hline
        \end{tabular}
     \end{subtable}
       \caption{Routing Table and Leafset of 456B in Chord Network}   
     \label{tab:chord Table 2}
\end{table}
\begin{table}[h]
    \begin{subtable}[h]{0.22\textwidth}
        \centering
        \begin{tabular}{|c | c |}
        \hline
            \multicolumn{2}{|c|}{Routing Table} \\\hline 
                Value of \lq${i}$\rq & NodeID \\\hline
                0 & 4E56  \\\hline
                1 & 4E56 \\\hline
                2 & 4E56 \\\hline
                3 & 4E56 \\\hline
                4 & 4E56 \\\hline
                5 & \textbf{4ECD} \\\hline
                6 & 6754 \\\hline
                7 & 0156 \\\hline
       \end{tabular}
    \end{subtable}
    \hfill
    \begin{subtable}[h]{0.26\textwidth}
        \centering
        \begin{tabular}{|l | l |}
        \hline
            \multicolumn{2}{|c|}{Leafset} \\\hline     Predecessor(P) & 456B  \\\hline
                Successor(S) & 4E56  \\\hline
        \end{tabular}
     \end{subtable}
       \caption{Routing Table and Leafset of 4ABC in Chord Network}   
     \label{tab:chord Table 3}
\end{table}
\begin{table}[h]
    \begin{subtable}[h]{0.22\textwidth}
        \centering
        \begin{tabular}{|c | c |}
        \hline
            \multicolumn{2}{|c|}{Routing Table} \\\hline 
               Value of \lq${i}$\rq & NodeID \\\hline
                0 & 4EF7  \\\hline
                1 & 4EF7 \\\hline
                2 & 4EF7 \\\hline
                3 & 6754 \\\hline
                4 & 6754 \\\hline
                5 & 6754 \\\hline
                6 & 6754 \\\hline
                7 & 0156 \\\hline
       \end{tabular}
    \end{subtable}
    \hfill
    \begin{subtable}[h]{0.26\textwidth}
        \centering
        \begin{tabular}{|c | c |}
        \hline
            \multicolumn{2}{|c|}{Leafset} \\\hline 
                Predecessor(P) & 4EAB  \\\hline
                Successor(S) & \textbf{4EF7}  \\\hline
        \end{tabular}
     \end{subtable}
       \caption{Routing Table and Leafset of 4ECD in Chord Network}   
     \label{tab:chord Table 4}
\end{table}
\hspace{5mm}
Table \ref{tab:chord Table 2} shows routing table and leafset of 456B. Using routing table, 456B will pass the query 4EFA to 4ABC. 4ABC will handover it to 4ECD. 4ECD to 4EF7 and 4EF7 will handover to 4EFB, which is the root node.

\item {\large{\textbf{{Pastry}}}}:
Here, nodeIDs and hashIDs will be considered as one single digit.
We have a similar metric for distance as that of a chord from Eqn(\ref{eq:GD2}) except now distance is symmetric and is equal to
$$\min_{R} \left(\left\{\left(R-H+2^d\right) \bmod 2^d\right\},\left\{\left(H-R+2^d\right) \bmod 2^d\right\}\right).$$
\hspace{5mm} If for two nodes, this value is same then node with cyclically lower value is used as root node. If we use the metric used in chord at all nodes, then also the algorithm will function correctly. Let us consider 4-digit hexadecimal nodeIDs. Each node will have a routing table with size, $15$ rows * $4$ columns, along with a leaf set of $k/2$ predecessors and $k/2$ successors. Assuming $k=4$ ($k=2m, m \geq 1$), there will be $2$ successors as well as predecessor.\par
\hspace{5mm} When a query for hashID is received by a node, it will try to find longest prefix match in the routing table. If there are more than one match, it will choose the numerically closer node as the next hop. If the current node is the longest prefix match and is also numerically closest among all options of longest prefix matches, the node will use leafset to forward the query to numerically closest node or the lower node if the two nodes are numerically closest to hashID. \par 
\hspace{5mm}NodeID 03A6 will have routing table as shown in figure \ref{fig:1} and leaf set as 0359, 0379 as P, and 03A9, 03AF as S as shown in table \ref{tab:Pastry leafset}. Table \ref{tab:Pastry routing table of 03A6} shows routing table of 03A6.
\par
\hspace{5mm} 03A6 will begin searching first in the routing table by prefix. 456B is the node in the routing table of 03A6, which has the longest prefix matching. So, it will transfer the query to 456B.\par
\hspace{5mm} Node 456B will hand it over to 4EAB, and 4EAB will transfer to 4EFC. 4EFC will transfer the query to 4EFB which is finally the root node.
\begin{table}[h!]
\centering
\caption{Routing Table of 03A6 in Pastry Network}
\begin{tabular}{|c|c|c|c|c|}
\hline
0 & 03A6 &  &  & \\ \hline
1 &  & 0156 &  & \\ \hline
2 & 25AB &  &  & \\ \hline
3 &  &03A6  &  & \\ \hline
4 & 456B  &04A5  &  & \\ \hline
5 &  &  &0359  & \\ \hline
6 & 6754 &  &  &03A6 \\ \hline
7 &  &  &0379  & \\ \hline
8 &  &  &  & \\ \hline
9 &  &  &  &03A9 \\ \hline
A &  &  &03A6  & \\ \hline
B &  &  &  & \\ \hline
C &  &  &  & \\ \hline
D &  &  &  & \\ \hline
E &  &  &  & \\ \hline
F &  &  &  &03AF \\ \hline
\end{tabular}
\label{tab:Pastry routing table of 03A6}
\end{table}

\begin{table}[h!]
\centering
\caption{Leafset of 03A6 in Pastry Network}
\begin{tabular}{|c|c|}
\hline
\multicolumn{2}{|c|}{Leafset}
\\\hline 
Predecessor & Successor \\ \hline
0359 & 03A9 \\ \hline
0379 & 03AF \\ \hline
\end{tabular}
\label{tab:Pastry leafset}
\end{table}
\end{enumerate}

\section{Different Routing Table sizes}
In any peer-to-peer system, the same distance metric for DHT should be used by all the nodes, i.e, the root node for a key will be the same in view of all the nodes. However, users can have different routing table sizes depending on their memory capacity.
$$
\textit{ Memory capacity } \propto \textit { Size of Routing Table}.
$$
\hspace{5mm} But there will be a trade-off between size of routing table and number of hop counts to reach the root node.
$$
\textit { Size of Routing Table } \propto(1 / \textit { number of hops}).
$$
\hspace{5mm} As the number of nodes in the network grows, the routing table entries increase, allowing for more efficient routing of messages. At the same time, the average number of hop count decreases, as the network becomes more connected and the distances between nodes decrease. In general, as the network becomes larger, the number of routing table entries increase, while the number of hop counts decrease.\par
\hspace{5mm} Consider a network, in which Tapestry is implemented to find out the root node for a key. Suppose in a system, we have a 40-bit hexadecimal nodeID (160-bit nodeID). The maximum size of a routing table is $15*40$ and the minimum size is $2*40$. Maximum routing table size has 600 entries and minimum routing table size has 80 entries. Peers can have different routing table with varying number of rows ranging from 2 to 15 i.e, size of routing table is $(X*40)$ where the value of $X$ lies between $2$ and $15$.\par
\hspace{5mm} The hop count in Tapestry is proportional to the logarithm of the size of the network. The number of columns in the routing table determines the size of the routing table, which in turn affects the hop count for routing messages in the network.

\section{Conclusion}
We presented a generalised formula for distance metric that expresses a variety of algorithms, including chord, pastry, tapestry, and kademlia. This study also suggests that the number of hops and memory capacity are contradicting objectives. Users can select one over the other.
\\[\baselineskip]\textbf{\large{Declarations}}
\\[\baselineskip]\textbf{Author contributions} Rashmi Kushwaha wrote the whole manuscript and worked out all the examples, Shreyas Kulkarni have verified all the formulas and diagrams and Yatindra Nath Singh gave the initial idea, reviewed and edited the manuscript, and verified all the examples.
\\[\baselineskip]\textbf{Funding} Rashmi Kushwaha and Shreyas Kulkarni received the fellowship from the Government of India.
\\[\baselineskip]\textbf{Data availability} Not applicable
\\[\baselineskip]\textbf{Ethics approval} No ethical clearance needed for this work.
\\[\baselineskip]\textbf{Consent to publish} Yes
\\[\baselineskip]\textbf{Conflict of Interest} The authors declare no conflict of Interest.

\vskip -2\baselineskip 

\begin{IEEEbiography}[{\includegraphics[width=1in,height=1.25in,clip,keepaspectratio]{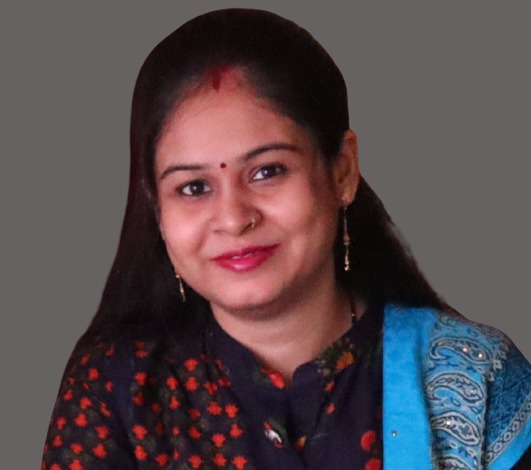}}]{Rashmi Kushwaha} She was born in Kanpur, India. She is currently pursuing Ph.D. in Electrical Engineering at IIT Kanpur. She holds an M.Tech in Digital Systems from MNNIT Prayagraj (formerly MNNIT Allahabad) and a B.Tech in Electronics Engineering from AKTU (formerly UPTU). Her research interests encompass Peer-to-Peer Networks, game theory applications, distributed networks, Internet of Things (IoT), smart networks, Blockchain, and cryptocurrency. With a proven track record in academia, she stands as a promising scholar poised to make significant contributions to the field of Electrical and Electronics Engineering.
\end{IEEEbiography}
\vskip -2\baselineskip plus -1fil
\begin{IEEEbiography}[{\includegraphics[width=1in,height=1.25in,clip,keepaspectratio]{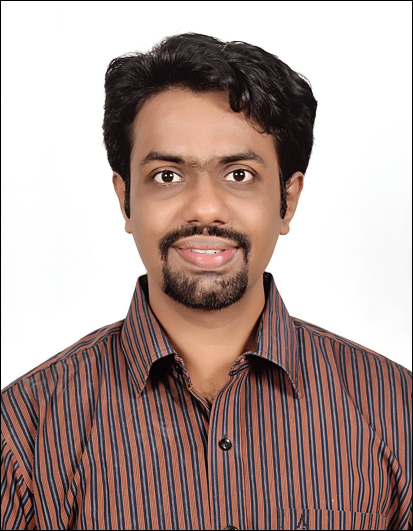}}]{Shreyas Kulkarni}He received his B.E degree from MIT Pune University in 2018 and completed his M.S degree in 2023 at Indian Institute of Technology Kanpur, specializing in signal processing and communication networks. With a robust background, he has proven expertise in software development, automotive security, and diagnostics. Currently based in Japan, Shreyas is making significant contributions to the telecommunications industry. His professional pursuits align with his keen interests in networking, cybersecurity, and distributed systems, where he seeks innovative approaches and collaborations to further advance these fields.
\end{IEEEbiography}
\vskip -2\baselineskip plus -1fil
\begin{IEEEbiography}[{\includegraphics[width=1in,height=1.25in,clip,keepaspectratio]{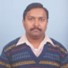}}]{Yatindra Nath Singh}He was born in Delhi, India. He did his B.Tech Electrical Engineering from REC Hamirpur (Now NIT Hamirpur), and M.Tech in Optoelectronics and Optical Communications from IIT Delhi. He was awarded Ph.D for his work on optical amplifier placement problem in all-optical broadcast networks in 1997 by IIT Delhi. In July 1997, he joined EE Department, IIT Kanpur. He was given AICTE young teacher award in 2003. Currently, he is working as professor. He was Head, Computer Center and Associate Dean of Digital Infrastructure from October 2015 to November 2017, Dean of Infrastructure and Planning IIT Kanpur from August 2017 to July 2020. He is fellow of IETE, fellow and member BOG of ICEIT, senior member of IEEE, and member ISOC. He has interests in telecommunications' networks specially in peer to peer networks, optical networks, switching systems, mobile communications, distributed software system design. He has supervised 21 Ph.D and more than 162 M.Tech/MSR theses so far.
He started a Technology Cooperative Startup www.yrrna.com in 2023, with his doctoral student for doing research and development for monetization.
He has been awarded three patents, and have published many journal and conference research publications. He has also written lecture notes on Digital Switching which are distributed as open access content through content repository of IIT Kanpur. He has also been involved in opensource software development. He had started Brihaspati (brihaspati.sourceforge.net) initiative, an opesource learning management system – current version Brihaspati3, BrihaspatiSync – a live lecture delivery system over Internet, BGAS – general accounting systems for academic institutes. Currently, he is pursuing research related to secure private encrypted peer-to-peer system named Brihaspati-4: a serverless LMS, telephony, messaging, live streaming system, search engine, user cloud, and collaborative compute system.
\end{IEEEbiography}
\vskip -2\baselineskip plus -1fil


\begin{thebibliography}{00}
\bibitem{b1}Napster, “Peer-to-peer music focused online services,” [Available Online] http://www.us.napster.com.

\bibitem{b2}Gnutella, “Gnutella protocol specification,” [Available Online] https://www.
rfc-gnutella.sourceforge.net/index.html.

\bibitem{b3}Freenet, “Peer-to-peer platform for communication and publication,” [Available Online] https://freenetproject.org/index.html.
\bibitem{b4}I. Stoica, R. Morris, D. Liben-Nowell, D. R. Karger, M. F. Kaashoek, F. Dabek, H. Balakrishnan, “Chord: a scalable peer-to-peer lookup protocol for internet applications," \textit{IEEE/ACM Transactions on Networking}, vol. 11, no. 1, pp. 17-32, 2003.

\bibitem{b5}B. Y. Zhao, L. Huang, J. Stribling, S. C. Rhea, A. D. Joseph, J. D. Kubiatowicz, “Tapestry: A resilient global-scale overlay for service deployment," \textit{IEEE Journal on selected areas in communications}, vol. 22, no. 1, pp. 41-53, 2004.

\bibitem{b6}A. Rowstron, P. Druschel, “Pastry: Scalable, decentralized object location, and routing for large-scale peer-to-peer systems," In \textit{Middleware 2001: IFIP/ACM International Conference on Distributed Systems Platforms Heidelberg, Germany, November 12–16, 2001 Proceedings 2}, pp. 329-350. Springer Berlin Heidelberg, 2001.

\bibitem{b7}P. Maymounkov, D. Mazieres, “Kademlia: A peer-to-peer information system based on the xor metric," In \textit{Peer-to-Peer Systems: First InternationalWorkshop, IPTPS 2002 Cambridge, MA, USA, March 7–8, 2002 Revised Papers}, pp. 53-65. Berlin, Heidelberg: Springer Berlin Heidelberg, 2002.

\bibitem{b8}J. Li, “Routing tradeoffs in dynamic peer-to-peer networks," PhD diss., Massachusetts Institute of Technology, 2005.

\bibitem{b9}L. KABRE, T. Tiendrebeogo, “Comparative study of Can, Pastry, Kademlia and Chord DHTs," In \textit{International Journal of Peer to Peer Networks (IJP2P)} Vol 12, No.1/2/3, August 2021.

\bibitem{b10}S. Ratnasamy, P. Francis, M. Handley, R. Karp, S. Shenker, “A scalable content-addressable network," In \textit{Proceedings of the 2001 conference on Applications, technologies, architectures, and protocols for computer communications}, pp. 161-172, 2001.

\bibitem{b11}A. Betts, L. Liu, Z. Li, N. Antonopoulos, “A critical comparative evaluation on DHT-based peer-to-peer search algorithms," \textit{International Journal of Embedded Systems}, vol. 6, no. 2-3, pp. 250-256, 2014.

\bibitem{b12}P. Flocchini, A. Nayak, M. Xie, “Hybrid-chord: A peer-to-peer system based on chord," In \textit{Distributed Computing and Internet Technology: First International Conference, ICDCIT 2004, Bhubaneswar, India, December 22-24, 2004, Proceedings 1}, pp. 194-203, Springer Berlin Heidelberg, 2005.

\bibitem{b13}M. F. Kaashoek, D. R. Karger, “Koorde: A simple degree-optimal distributed hash table," In \textit{Peer-to-Peer Systems II: Second International Workshop, IPTPS 2003, Berkeley, CA, USA, February 21-22, 2003}, Revised Papers 2, pp. 98-107, Springer Berlin Heidelberg, 2003.

\bibitem{b14}F. Chao, H. Zhang, X. Du, C. Zhang, “Improvement of structured P2P routing algorithm based on NN-CHORD," In \textit{2011 7th International Conference on Wireless Communications, Networking and Mobile Computing}, pp. 1-5, IEEE, 2011.

\bibitem{b15}Z. Guo, S. Yang, H. Yang, “P4P Pastry: A novel P4P-based Pastry routing algorithm in peer to peer network," In \textit{2010 2nd IEEE International Conference on Information Management and Engineering}, pp. 209-213, IEEE, 2010.

\end{thebibliography}
\end{document}